%% file: quartification2.tex
\documentclass [12pt] {article}
\usepackage{axodraw}
\usepackage{color}

\catcode`@=12 \topmargin -0.5in \evensidemargin 0.0in \oddsidemargin 0.0in
\begin{document}
\thispagestyle{empty}
\begin{flushright} hep-ph/0307380\\UCRHEP-T358\\
OSU-HEP-03-09\\ILL-(TH)-03-04\\NSF-KITP-03-56\\July 2003
\
\end{flushright}
\vspace{0.5in}
\begin{center}
{\LARGE \bf Quark-Lepton Quartification\\} \vspace{0.8in} {\bf K.S. Babu$^1$,
Ernest Ma$^{2,4}$, and S. Willenbrock$^{3,4}$\\} \vspace{0.2in} {\sl $^1$
Physics Department, Oklahoma State University, Stillwater, Oklahoma 74078,
USA\\} {\sl $^2$ Physics Department, University of California, Riverside,
California 92521, USA\\} {\sl $^3$ Department of Physics, University of
Illinois at Urbana-Champaign,\\} {\sl 1110 West Green Street, Urbana, Illinois
61801, USA\\} {\sl $^4$ Kavli Institute for Theoretical Physics, University of
California,\\} {\sl Santa Barbara, California 93106, USA\\} \vspace{1.2in}
\end{center}

\begin{abstract}
We propose that quarks and leptons are interchangeable entities in the
high-energy limit.  This naturally results in the extension of ${[SU(3)]}^3$
trinification to ${[SU(3)]}^4$ quartification.  In addition to the unbroken
color $SU(3)_q$ of quarks, there is now also a color $SU(3)_l$ of leptons
which reduces to an unbroken $SU(2)_l$. We discuss the natural occurrence of
$SU(2)_l$ doublets at the TeV energy scale, which leads remarkably to the
unification of all gauge couplings without supersymmetry. Proton decay occurs
through the exchange of scalar bosons, with a lifetime in the range $10^{34} -
10^{36}$ years.
\end{abstract}

\newpage
\baselineskip 24pt

Leptons are different from quarks in two important ways.  (1) Quarks come in
three colors and they interact as triplets under the color $SU(3)$ gauge group,
whereas leptons are singlets and do not interact at all in this respect. (2)
Quarks have fractional electric charges, whereas leptons have integral
charges.  These differences are obvious but perhaps they are also
superficial.  By that we mean the possibility that leptons and quarks are
actually very much alike in the high-energy limit, and they only appear to be
different at low energies.

The implementation of this idea turns out to be very simple and natural.  We
merely accept the existence of a separate lepton color $SU(3)_l$ gauge symmetry
\cite{Foot:dw}, which is broken at a high scale to $SU(2)_l$.  Two of the three
components of this lepton color triplet are confined by the residual $SU(2)_l$
gauge interactions, but the third is unconfined, and that is what we observe
at low energies.  At the same time, the partial breaking of the lepton color
$SU(3)_l$ allows us to understand the electric charge differences between
quarks and the observed leptons.

The minimum gauge group containing leptonic color is $SU(3)_q \times SU(3)_l
\times SU(2)_L \times U(1)$. This is very interesting by itself and has been
thoroughly discussed \cite{Foot:dw,Foot:fk}.  Partial unification of $SU(3)_q
\times SU(3)_l \times SU(3)_L \times SU(3)_R$ has also been considered
\cite{jv}.  Here we find the remarkable new result that the natural occurrence
of exotic $SU(2)_l$ doublet fermions at the TeV scale automatically leads to
the complete unification of all gauge couplings at around $10^{11}$ GeV
without supersymmetry.  We call this $[SU(3)]^4$ ``quartification''.

Quarks and leptons interact with each other through a common set of gauge
bosons, i.e. the photon and the $W^\pm$ and $Z^0$ bosons.  In addition, quarks
interact among themselves through the $SU(3)_c$ gluons.  This pattern is
somewhat asymmetric, but it may only be so because it is the low-energy
remnant of a much more symmetric theory at high energies. An intriguing
possibility is trinification based on the gauge group $SU(3)_c \times SU(3)_L
\times SU(3)_R$ \cite{deRujula,Babu:gi,Lazarides:sn,Willenbrock:2003ca}, under
which the quarks and leptons belong to the $(3,3^*,1)$, $(3^*,1,3)$, and
$(1,3,3^*)$ representations, as shown in Fig.~1.  We adopt the convention that
all fermion fields are left-handed, with their right-handed counterparts
denoted by the corresponding (left-handed) charge-conjugate fields.

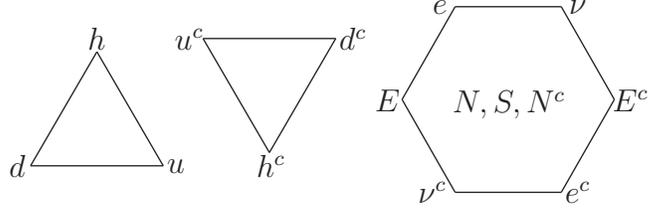
\begin{figure}[htb]
\begin{center}
\begin{picture}(270,60)(0,0)
\Line(10,10)(60,10) \Line(10,10)(35,53) \Line(35,53)(60,10) \Text(35,58)[]{$h$}
\Text(5,10)[]{$d$} \Text(65,10)[]{$u$}

\Line(75,58)(125,58) \Line(75,58)(100,15) \Line(100,15)(125,58)
\Text(70,58)[]{$u^c$} \Text(132,58)[]{$d^c$} \Text(101,10)[]{$h^c$}

\Line(170,0)(210,0) \Line(150,35)(170,0) \Line(150,35)(170,70)
\Line(170,70)(210,70) \Line(210,70)(230,35) \Line(210,0)(230,35)
\Text(165,70)[]{$e$} \Text(217,70)[]{$\nu$} \Text(162,0)[]{$\nu^c$}
\Text(217,0)[]{$e^c$} \Text(145,35)[]{$E$} \Text(237,35)[]{$E^c$}
\Text(191,33)[]{$N,S,N^c$}

\end{picture}
\end{center}
\caption{Pictorial representation of quarks and leptons in $[SU(3)]^3$
trinification. The (implicit) $x,y$ axes in the first diagram are
$I_{3L},Y_L$; in the second diagram, $I_{3R},Y_R$; and in the third diagram,
$I_{3L}+I_{3R},Y_L+Y_R$.}
\end{figure}

The electric charge $Q$ is given by
\begin{equation}
Q = I_{3L} - {Y_L \over 2} + I_{3R} - {Y_R \over 2} = I_{3L} + {Y \over 2}.
\end{equation}
Hence the exotic fermion $h(h^c)$ has charge $\mp 1/3$, $E(E^c)$ has charge
$\mp 1$, and $N,N^c,S$ are neutral.  In matrix notation, the leptons and
antileptons are contained in
\begin{equation}
\lambda \sim \pmatrix {N & E^c & \nu \cr E & N^c & e \cr \nu^c & e^c & S},
\end{equation}
where $I_{3L} = (1/2, -1/2, 0)$ and $Y_L = (1/3, 1/3, -2/3)$ for the rows,
and  $I_{3R} = (-1/2, 1/2, 0)$ and $Y_R = (-1/3, -1/3, 2/3)$ for the columns.
The quarks are given by
\begin{equation}
q \sim \pmatrix {d & u & h \cr d & u & h \cr d & u & h}, ~~~ q^c \sim \pmatrix
{d^c & d^c & d^c \cr u^c & u^c & u^c \cr h^c & h^c & h^c},
\end{equation}
where $I_{3L} = (-1/2, 1/2, 0)$ for the columns in $q$ and $I_{3R} = (1/2,
-1/2, 0)$ for the rows in $q^c$.  The addition of two scalar multiplets
\begin{equation}
\phi_a \sim (1,3,3^*)~~(a=1,2)
\end{equation}
allows the Yukawa terms $Tr(q^c q \phi_a)$ as well as $\epsilon_{ijk}
\epsilon_{mnp} \lambda^{im} \lambda^{jn} \phi_a^{kp}$, thus providing all
fermions with appropriate masses and mixings. (With only one such $\phi$
field, the $up$ and $down$ quark mass matrices would be proportional to each
other and all charged-current mixing angles would be zero.) Using a ``moose''
\cite{Georgi:1985hf} or ``quiver'' \cite{Douglas:1996sw} diagram, this model
may be depicted as in Fig.~2.

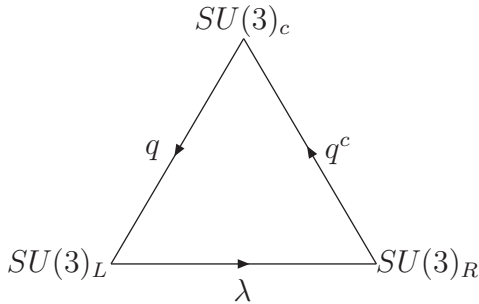
\begin{figure}[htb]
\begin{center}
\begin{picture}(300,100)(0,0)
\ArrowLine(65,10)(165,10) \ArrowLine(115,95)(65,10) \ArrowLine(165,10)(115,95)
\Text(115,102)[]{$SU(3)_c$} \Text(45,10)[]{$SU(3)_L$}
\Text(185,10)[]{$SU(3)_R$} \Text(81,53)[]{$q$} \Text(151,53)[]{$q^c$}
\Text(115,0)[]{$\lambda$}

\end{picture}
\end{center}
\caption{Moose diagram of $[SU(3)]^3$ trinification.}
\end{figure}

It is clear that quarks and leptons are still dissimilar in the $[SU(3)]^3$
model.  To achieve complete quark--lepton symmetry, we make use of the notion
\cite{Foot:dw} that leptons may also come in three colors under a separate
color $SU(3)_l$ gauge group, which is then broken to $SU(2)_l$, allowing only
one component, i.e. the observed lepton, to be light and unconfined.  In that
case, we have a natural $[SU(3)]^4$ model with the moose diagram as shown in
Fig.~3, which is clearly totally symmetric with respect to the interchange of
quarks and leptons.  [In contrast, the extension of trinification to include
chiral color \cite{Frampton:1987dn}, i.e. $SU(3)_c \to SU(3)_{cL} \times
SU(3)_{cR}$, would not be.]

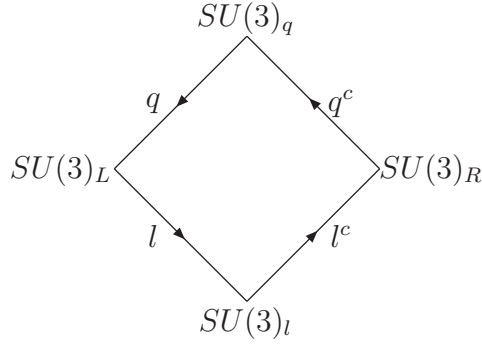
\begin{figure}[htb]
\begin{center}
\begin{picture}(300,110)(0,0)
\ArrowLine(120,5)(170,55) \ArrowLine(170,55)(120,105) \ArrowLine(70,55)(120,5)
\ArrowLine(120,105)(70,55) \Text(120,112)[]{$SU(3)_q$}
\Text(120,-3)[]{$SU(3)_l$} \Text(50,55)[]{$SU(3)_L$} \Text(190,55)[]{$SU(3)_R$}
\Text(85,80)[]{$q$} \Text(156,80)[]{$q^c$} \Text(85,30)[]{$l$}
\Text(156,30)[]{$l^c$}

\end{picture}
\end{center}
\caption{Moose diagram of $[SU(3)]^4$ quartification.}
\end{figure}

Under $SU(3)_q \times SU(3)_L \times SU(3)_l \times SU(3)_R$, we then have
\begin{eqnarray}
q \sim (3,3^*,1,1), &~& q^c \sim (3^*,1,1,3), \\
l \sim (1,3,3^*,1), &~& l^c \sim (1,1,3,3^*).
\end{eqnarray}
Although chiral, this assignment is free of all triangle anomalies. Whereas
$q$ and $q^c$ may be depicted in the same way as in Fig.~1 and represented as
in Eq.~(3), $l$ and $l^c$ replace $\lambda$ of Eq.~(2).  The analog of Eq.~(1)
becomes
\begin{equation}
Q = I_{3L} - {Y_L \over 2} + I_{3R} - {Y_R \over 2} - {Y_l \over 2}.
\end{equation}
Here $Y_l$ takes the same values as $Y_L$ or $Y_R$ depending of course on
whether it is part of a triplet or antitriplet.  The matrix representations of
$l$ and $l^c$ are
\begin{equation}
l \sim \pmatrix {x_1 & x_2 & \nu \cr y_1 & y_2 & e \cr z_1 & z_2 & N}, ~~~ l^c
\sim \pmatrix {x_1^c & y_1^c & z_1^c \cr x_2^c & y_2^c & z_2^c \cr \nu^c & e^c
& N^c},
\end{equation}
where the columns of $l$ have $Y_l = (-1/3,-1/3,2/3)$, and the rows of $l^c$
have $Y_l$ = (1/3,1/3, --2/3). Using Eq.~(7), we find $N$ and $N^c$ to be
neutral, and the exotic $SU(2)_l$ doublet leptons $(x,y,z)$ to have charges
$(1/2,-1/2,1/2)$ and $(x^c,y^c,z^c)$ to have charges $(-1/2,1/2,-1/2)$
respectively.  Because of their half integral charges, we call the $SU(2)_l$
doublets ``hemions''.

To obtain the usual lepton and quark masses, we need the analogs of the scalar
multiplets of Eq.~(4), i.e.
\begin{equation}
\phi_a \sim (1,3^*,1,3)~~(a=1,2),
\end{equation}
which allow the Yukawa terms $Tr (l l^c \phi_a)$ and $Tr (q^c q
\phi_a^\dagger)$. As in the case of trinification, two such $\phi$ fields are
necessary to generate  realistic quark masses and nonzero mixing angles. In
the basis $I_{3L} = (-1/2,1/2,0)$ for the columns and $I_{3R} = (1/2,-1/2,0)$
for the rows, we note that $(\phi_a)_{ij}$ has the same charge assignments as
$\lambda$ of Eq.~(2).  Two other scalar multiplets
\begin{equation}
\phi_L \sim (1,3,3^*,1) \sim l, ~~~ \phi_R \sim (1,1,3,3^*) \sim l^c,
\end{equation}
are also assumed, by which $SU(3)_l$ may be broken down to $SU(2)_l$. The
Yukawa terms $\epsilon_{ijk} \epsilon_{mnp} l^{im} l^{jn} \phi_L^{kp}$ and
$\epsilon_{ijk} \epsilon_{mnp} (l^c)^{im} (l^c)^{jn} \phi_R^{kp}$ are {\it a
priori} possible, but we forbid them by a discrete symmetry to be discussed
later.

Consider the neutral components of $\phi_L$, $\phi_R$, and $\phi$ that are
also singlets under $SU(3)_q \times SU(2)_l \times SU(2)_L \times U(1)_Y$.
There are five such fields: $\phi_L^N$, $\phi_R^{N^c}$, $\phi_R^{\nu^c}$,
$\phi_{13}$, and $\phi_{33}$. [We denote the components of $\phi_L,\phi_R$ by
the corresponding components of $l,l^c$.]  They are connected by the trilinear
scalar coupling $\phi_L \phi_R \phi$ in the combination $\phi_L^N
\phi_R^{\nu^c} \phi_{13} + \phi_L^N \phi_R^{N^c} \phi_{33}$. This means that
their vacuum expectation values are all naturally of the same order of
magnitude, in which case $[SU(3)]^4$ breaks down completely to $SU(3)_q \times
SU(2)_l \times SU(2)_L \times U(1)_Y$ at this one scale.  Since $hd^c$ couples
to $\phi_{13}^*$, $hh^c$ to $\phi_{33}^*$, $N\nu^c$ to $\phi_{13}$, $NN^c$ to
$\phi_{33}$, $zx^c$ to $\phi_{13}$, $zz^c$ to $\phi_{33}$, heavy Dirac masses
are obtained for $h,N,z$ with linear combinations of $d^c$ and $h^c$, $\nu^c$
and $N^c$, $x^c$ and $z^c$ respectively.  Redefining the orthogonal
combinations of the latter to be simply $d^c, \nu^c, x^c$, the massless
fermions of our model consists of those of the minimal standard model plus
$\nu^c$ and the hemions $(x,y), x^c, y^c$.  This particle spectrum turns out
to be exactly what is needed for the unification of all the gauge couplings at
around $10^{11}$ GeV as shown below.

Consider the possibility that the four $SU(3)$ gauge couplings are equal at
some high scale.  Instead of embedding $[SU(3)]^4$ in a single simple group as
in ordinary grand unification, we invoke a cyclic $Z_4$ symmetry which rotates
the four gauge groups in the manner indicated by the moose diagram of Fig.~3.
The fermionic spectrum is already compatible with such a $Z_4$ symmetry with
$q \to l \to l^c \to q^c \to q$.  The Higgs sector is enlarged to accommodate
$\phi_L \to \phi_R \to \phi_R' \to \phi_L' \to \phi_L$, whereas $\phi_1$ and
$\phi_2$ (renamed $\phi_3^\dagger$) are contained in the $Z_4-$invariant Yukawa
Lagrangian
\begin{eqnarray}
{\cal L}_Y &=& Y_1 Tr(l l^c \phi_1 + l^c q^c \phi_2 + q^c q \phi_3 + q l
\phi_4) \nonumber \\ &+& Y_2 Tr(l l^c \phi_3^\dagger + l^c q^c \phi_4^\dagger
+ q^c q \phi_1^\dagger + q l \phi_2^\dagger),
\end{eqnarray}
with the new $\phi_2$ transforming as $(3,1,3^*,1)$, etc. We then have ${\cal
M}_l = Y_1 \langle (\phi_1)_{22} \rangle + Y_2 \langle (\phi_3)_{22}^*
\rangle$, ${\cal M}_u = Y_1 \langle (\phi_3)_{22} \rangle + Y_2 \langle
(\phi_1)_{22}^* \rangle$, and ${\cal M}_d = Y_1 \langle (\phi_3)_{11} \rangle +
Y_2 \langle (\phi_1)_{11}^* \rangle$, if the $d^c-h^c$ mixing described above
is ignored.

The standard model (without supersymmetry) fails to yield coupling
unification, assuming the standard  $SU(5)$ relation $\sin^2\theta_W=3/8$ at
the unification scale.  This same relation is true in the $[SU(3)]^3$
trinification model.  However, in the $[SU(3)]^4$ quartification model,
$\sin^2\theta_W=1/3$ at the unification scale, which follows from the
embedding of electric charge, i.e. Eq.~(7).  This exacerbates the failure of
the standard model regarding coupling unification.  On the other hand, we have
relatively light $(x,y), x^c, y^c$ hemions in this model and they will affect
the evolution of all the gauge couplings except that of $SU(3)_q$.

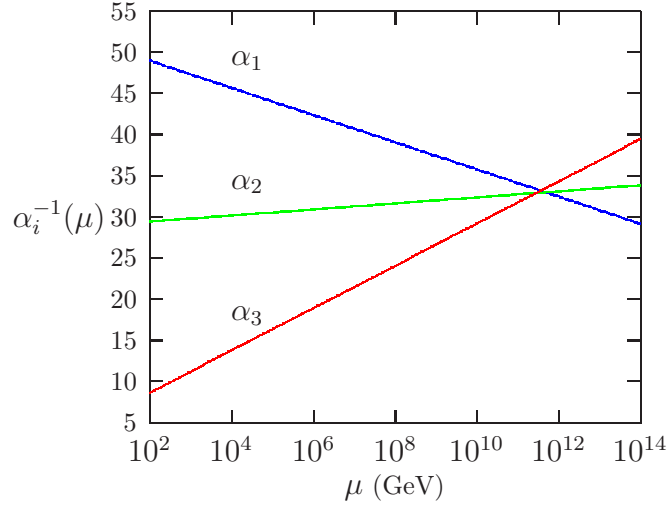
\begin{figure}
\begin{center}
\input{quartfigsmall}
\end{center}
\caption{Leading-order evolution of the gauge couplings from their low-energy
values to the unification scale in a simple $[SU(3)]^4$ quartification model
with no intermediate scales.}\label{quartfigsmall}
\end{figure}

The renormalization-group evolution of the gauge couplings is dictated at
leading order by
\begin{equation}
\frac{1}{\alpha_i(\mu)}-\frac{1}{\alpha_i(\mu')}=\frac{b_i}{2\pi}
\ln\left(\frac{\mu'}{\mu}\right)\;,\label{rng}
\end{equation}
where $b_n$ are the one-loop beta-function coefficients,
\begin{eqnarray}
&&b_3 = -11 + \frac{4}{3}N_g, \\
&&b_2 = -\frac{22}{3} +2N_g + \frac{1}{6}N_H, \\
&&b_1 = \frac{13}{9}N_g + \frac{1}{12}N_H,
\end{eqnarray}
($N_g=3$ is the number of generations) which include the contributions of the
weak-scale $SU(2)_l$ doublet hemions [$(x,y)$ is an $SU(2)_L$ doublet with
$Y=0$; $x^c$ and $y^c$ are $SU(2)_L$ singlets with $Y=\mp1$] and $N_H$ Higgs
doublets with $Y=\pm 1$.  The initial values of the gauge couplings are
\begin{eqnarray}
&&\alpha_3(M_Z)=0.117, \\
&&\alpha_2(M_Z)=(\sqrt 2/\pi)G_FM_W^2 = 0.034, \\
&&\alpha_1(M_Z) = 2\alpha_2(M_Z)\tan^2\theta_W = 0.0204,
\end{eqnarray}
where the factor of $2$ in the last relation [and the normalization of $b_1$
in Eq.~(15)] is determined by the embedding of $U(1)_Y$ in $[SU(3)]^4$.  We
show the evolution of the couplings from the weak scale up to very high scales
in Fig.~\ref{quartfigsmall}, using $N_H=2$. Remarkably, the couplings actually
meet, within the accuracy of the leading-order calculation, at $4 \times
10^{11}$ GeV.  Even more Higgs doublets can be accommodated if some hemions
are heavier than the weak scale.

The evolution of the unbroken $SU(2)_l$ coupling from the unification scale to
the weak scale is given by Eq.~(12), with $b_{2l}=-22/3+(4/3)N_g$, which
includes the contribution of the weak-scale hemions.  This yields
$\alpha_{2l}^{-1}(M_Z)\approx 21$, which is between the weak and strong
couplings.  Below the weak scale, the evolution is driven only by the
$SU(2)_l$ gauge bosons.  The coupling becomes strong at a scale $\Lambda_{2l}
\approx 1$ keV, five orders of magnitude less than $\Lambda_{QCD}$.

Let us look at the hemion masses in more detail. If the $l l \phi_L$ and $l^c
l^c \phi_R$ terms were not forbidden, then all hemions would be heavy at the
quartification scale.  As it is, they are forbidden by a discrete symmetry, so
certain hemions become massive only through $\phi_{11}$, $\phi_{22}$, and
$\phi_{31}$, with the same Yukawa couplings as the observed leptons, which is
not realistic.  Thus we consider the possible existence of the
nonrenormalizable terms such as $\epsilon_{ijk} \epsilon_{mnp} l^{im} l^{jn}
[(\phi_R \phi_1)^\dagger]^{kp}$ and $\epsilon_{ijk} \epsilon_{mnp}(l^c)^{im}
(l^c)^{jn} [(\phi_1 \phi_L)^\dagger]^{kp}$, which are suppressed presumably by
the Planck mass. Since $(10^{11} {\rm GeV})^2/(10^{19} {\rm GeV}) = 10^3$ GeV,
it is natural in our model to have hemion masses at the TeV scale.
Furthermore, the resulting mass terms ($x_1 y_2 - x_2 y_1$, etc.) are singlets
under $SU(2)_L \times U(1)_Y$ and thus have negligible contributions to the
oblique parameters in precision electroweak measurements and would not upset
the agreement of present data with the standard model.  As to what discrete
symmetry we should use, there are some options.  The key is to have $\phi_L$
different from $(\phi_R \phi_1)^\dagger$, etc. under this symmetry. An example
is $Z_4$, under which $(q,l,l^c,q^c)$ transform as $i$, $\phi$ as $-1$, and
$\phi_{L,R}$ as $+1$. [If we add a new $\phi$, then $Z_2$ may be used.]

In contrast to $SU(5)$ grand unification, baryon number is conserved here by
the gauge interactions of $[SU(3)]^4$ quartification.  However, it is violated
in the scalar and Yukawa sectors.  Consider for example the $Z_4-$invariant
dimension-five operator  $ll (\phi_R \phi_1)^\dagger + l^c l^c (\phi_R'
\phi_2)^\dagger + q^c q^c (\phi_L'\phi_3)^\dagger + q q (\phi_L
\phi_4)^\dagger$.  When $\phi_L^N$ acquires a vacuum expectation value, the
last term induces a diquark coupling $u d (\phi_4)_{33}^*$, with a strength of
order $\lambda' = 10^{11}/10^{19} = 10^{-8}$.  In addition, the top quark
Yukawa coupling from Eq.~(11) contains a leptoquark coupling of the same field
$(\phi_4)_{33}$ given by $\lambda_t b \nu_\tau (\phi_4)_{33}$.  The exchange
of $(\phi_4)_{33}$ then leads to proton decay, with a rate for the dominant
mode $\bar{\nu}_\tau K^+$ estimated to be
\begin{equation}
\Gamma(p \rightarrow \bar{\nu}_\tau K^+) \approx
 {(\lambda_t \lambda' \theta_{sb})^2
\over 8 \pi} {m_p^5 \over M^4_\phi } \sim [10^{35} y]^{-1},
\end{equation}
where $\lambda' = 10^{-8}$, $\lambda_t = 1$, $\theta_{sb} \simeq V_{cb} =
0.04$, and $M_\phi = 4 \times 10^{11}$ GeV.  If the Yukawa couplings that
induce the $x$ and $y$ hemion masses were not suppressed, this model would
have predicted too short a lifetime for the proton. Thus the TeV scale hemions
are needed both for gauge-coupling unification and for consistency with proton
decay.  Since Eq.~(19) has large numerical uncertainties, a more reasonable
estimate of the proton lifetime is in the range $10^{34}-10^{36}$ years.

Consider now the neutrino mass matrix.  It gets no contribution from the
antisymmetric $l l$ and $l^c l^c$ terms.  Thus it is of the purely Dirac form
derived from Eq.~(11). To obtain a small Majorana $m_\nu$, consider the
addition of a heavy fermion singlet $S$ per family, which is linked to $\nu$
by the Yukawa term $f Tr(l \phi_L^\dagger + l^c \phi_R^\dagger) S$.  For the
masses of $\nu^c, N, N^c, S$ all of order $10^{11}$ GeV and $f \langle
\phi_L^\nu \rangle \sim 10$ GeV, we get $m_\nu \sim 1$ eV, which is a very
reasonable mass scale for neutrino oscillations.  [An alternative is to add an
extra scalar multiplet $\xi_R \sim (1,1, 6^*, 6)$ with the Yukawa coupling
$l^c_{ij} l^c_{mn} \xi_R^{im,jn}$ to provide $N^c$ with a large Majorana mass.]

The $SU(2)_l$ gauge symmetry is unbroken.  In analogy to the gluons of
$SU(3)_q$, we call the 3 massless gauge bosons of $SU(2)_l$ ``stickons''. They
serve to confine the hemions into integrally charged particles.  They can also
stick together to form ``stickballs''.  In quantum chromodynamics, the
glueballs are very unstable because they are heavier than the mesons. Here the
corresponding ``mesons'' are much heavier than the stickballs. Since the
hemions are also electrically charged, a necessary decay mode of a stickball
is into 2 photons, but is suppressed by the mass of the hemion in the loop. At
the TeV scale, hemions may be accessible at future colliders.  Two hemions
will combine to form 3 kinds of particles: call them $\lambda^+, \lambda^0,
\lambda^-$.  Whereas $\lambda^0$ decays immediately into 2 stickballs,
$\lambda^\pm$ is somewhat more stable, because it has to decay into a $W^\pm$
and a stickball.

In conclusion, we have proposed a novel scenario for the unification of quarks
and leptons based on the gauge symmetry $SU(3)_q \times SU(3)_L \times SU(3)_l
\times SU(3)_R$.  Leptons become just like quarks in that there is also a
color $SU(3)_l$.  However only the subgroup $SU(2)_l$ is unbroken, and the
corresponding confined ``lepton'' doublets (which we call hemions) are either
very heavy or at the TeV scale, whereas the observed leptons are singlets and
unconfined.  We have discussed how this $[SU(3)]^4$ quartification is broken
at around  $10^{11}$ GeV to the gauge symmetries of the standard model plus
$SU(2)_l$, and how its particle content (that of the standard model plus
hemions at the TeV scale) leads remarkably well to the unification of the 3
known gauge couplings without supersymmetry and a proton lifetime in the range
$10^{34} - 10^{36}$ years. [Recall that in minimal nonsupersymmetric $SU(5)$,
gauge couplings do not unify and the predicted proton lifetime is too short.]

\section*{Acknowledgements}

This work was supported in part by the U.~S.~Department of Energy under Grant
Nos.~DE-FG03-94ER40837, DE-FG02-91ER40677, DE-FG03-98ER-41076,
DE-FG02-01ER-45684, by the National Science Foundation under Grant
No.~PHY99-07949 and an award from the Research Corporation.  E.~M. and S.~W.
thank the KITP (UC Santa Barbara) and the organizers of its 2003 Neutrino
Workshop for hospitality and the stimulating environment which fostered this
work.

\newpage
\bibliographystyle{unsrt}

\end{document}

%% file: quartfigsmall.tex
\setlength{\unitlength}{0.240900pt}
\ifx\plotpoint\undefined\newsavebox{\plotpoint}\fi
\begin{picture}(1012,809)(0,0)
\font\gnuplot=cmr10 at 10pt
\gnuplot
\sbox{\plotpoint}{\rule[-0.200pt]{0.400pt}{0.400pt}}%
\put(182.0,123.0){\rule[-0.200pt]{4.818pt}{0.400pt}}
\put(162,123){\makebox(0,0)[r]{5}}
\put(932.0,123.0){\rule[-0.200pt]{4.818pt}{0.400pt}}
\put(182.0,188.0){\rule[-0.200pt]{4.818pt}{0.400pt}}
\put(162,188){\makebox(0,0)[r]{10}}
\put(932.0,188.0){\rule[-0.200pt]{4.818pt}{0.400pt}}
\put(182.0,252.0){\rule[-0.200pt]{4.818pt}{0.400pt}}
\put(162,252){\makebox(0,0)[r]{15}}
\put(932.0,252.0){\rule[-0.200pt]{4.818pt}{0.400pt}}
\put(182.0,317.0){\rule[-0.200pt]{4.818pt}{0.400pt}}
\put(162,317){\makebox(0,0)[r]{20}}
\put(932.0,317.0){\rule[-0.200pt]{4.818pt}{0.400pt}}
\put(182.0,382.0){\rule[-0.200pt]{4.818pt}{0.400pt}}
\put(162,382){\makebox(0,0)[r]{25}}
\put(932.0,382.0){\rule[-0.200pt]{4.818pt}{0.400pt}}
\put(182.0,446.0){\rule[-0.200pt]{4.818pt}{0.400pt}}
\put(162,446){\makebox(0,0)[r]{30}}
\put(932.0,446.0){\rule[-0.200pt]{4.818pt}{0.400pt}}
\put(182.0,511.0){\rule[-0.200pt]{4.818pt}{0.400pt}}
\put(162,511){\makebox(0,0)[r]{35}}
\put(932.0,511.0){\rule[-0.200pt]{4.818pt}{0.400pt}}
\put(182.0,576.0){\rule[-0.200pt]{4.818pt}{0.400pt}}
\put(162,576){\makebox(0,0)[r]{40}}
\put(932.0,576.0){\rule[-0.200pt]{4.818pt}{0.400pt}}
\put(182.0,641.0){\rule[-0.200pt]{4.818pt}{0.400pt}}
\put(162,641){\makebox(0,0)[r]{45}}
\put(932.0,641.0){\rule[-0.200pt]{4.818pt}{0.400pt}}
\put(182.0,705.0){\rule[-0.200pt]{4.818pt}{0.400pt}}
\put(162,705){\makebox(0,0)[r]{50}}
\put(932.0,705.0){\rule[-0.200pt]{4.818pt}{0.400pt}}
\put(182.0,770.0){\rule[-0.200pt]{4.818pt}{0.400pt}}
\put(162,770){\makebox(0,0)[r]{55}}
\put(932.0,770.0){\rule[-0.200pt]{4.818pt}{0.400pt}}
\put(182.0,123.0){\rule[-0.200pt]{0.400pt}{4.818pt}}
\put(182,82){\makebox(0,0){$10^{2}$}}
\put(182.0,750.0){\rule[-0.200pt]{0.400pt}{4.818pt}}
\put(310.0,123.0){\rule[-0.200pt]{0.400pt}{4.818pt}}
\put(310,82){\makebox(0,0){$10^{4}$}}
\put(310.0,750.0){\rule[-0.200pt]{0.400pt}{4.818pt}}
\put(439.0,123.0){\rule[-0.200pt]{0.400pt}{4.818pt}}
\put(439,82){\makebox(0,0){$10^{6}$}}
\put(439.0,750.0){\rule[-0.200pt]{0.400pt}{4.818pt}}
\put(567.0,123.0){\rule[-0.200pt]{0.400pt}{4.818pt}}
\put(567,82){\makebox(0,0){$10^{8}$}}
\put(567.0,750.0){\rule[-0.200pt]{0.400pt}{4.818pt}}
\put(695.0,123.0){\rule[-0.200pt]{0.400pt}{4.818pt}}
\put(695,82){\makebox(0,0){$10^{10}$}}
\put(695.0,750.0){\rule[-0.200pt]{0.400pt}{4.818pt}}
\put(824.0,123.0){\rule[-0.200pt]{0.400pt}{4.818pt}}
\put(824,82){\makebox(0,0){$10^{12}$}}
\put(824.0,750.0){\rule[-0.200pt]{0.400pt}{4.818pt}}
\put(952.0,123.0){\rule[-0.200pt]{0.400pt}{4.818pt}}
\put(952,82){\makebox(0,0){$10^{14}$}}
\put(952.0,750.0){\rule[-0.200pt]{0.400pt}{4.818pt}}
\put(182.0,123.0){\rule[-0.200pt]{185.493pt}{0.400pt}}
\put(952.0,123.0){\rule[-0.200pt]{0.400pt}{155.862pt}}
\put(182.0,770.0){\rule[-0.200pt]{185.493pt}{0.400pt}}
\put(40,446){\makebox(0,0){$\alpha_{i}^{-1}(\mu)$}}
\put(567,21){\makebox(0,0){$\mu$ (GeV)}}
\put(310,692){\makebox(0,0)[l]{$\alpha_{1}$}}
\put(310,498){\makebox(0,0)[l]{$\alpha_{2}$}}
\put(310,291){\makebox(0,0)[l]{$\alpha_{3}$}}
\put(182.0,123.0){\rule[-0.200pt]{0.400pt}{155.862pt}}
{\color{blue}\put(182,692){\usebox{\plotpoint}}
\multiput(182.00,690.95)(1.579,-0.447){3}{\rule{1.167pt}{0.108pt}}
\multiput(182.00,691.17)(5.579,-3.000){2}{\rule{0.583pt}{0.400pt}}
\put(190,687.17){\rule{1.700pt}{0.400pt}}
\multiput(190.00,688.17)(4.472,-2.000){2}{\rule{0.850pt}{0.400pt}}
\multiput(198.00,685.95)(1.355,-0.447){3}{\rule{1.033pt}{0.108pt}}
\multiput(198.00,686.17)(4.855,-3.000){2}{\rule{0.517pt}{0.400pt}}
\multiput(205.00,682.95)(1.579,-0.447){3}{\rule{1.167pt}{0.108pt}}
\multiput(205.00,683.17)(5.579,-3.000){2}{\rule{0.583pt}{0.400pt}}
\put(213,679.17){\rule{1.700pt}{0.400pt}}
\multiput(213.00,680.17)(4.472,-2.000){2}{\rule{0.850pt}{0.400pt}}
\multiput(221.00,677.95)(1.579,-0.447){3}{\rule{1.167pt}{0.108pt}}
\multiput(221.00,678.17)(5.579,-3.000){2}{\rule{0.583pt}{0.400pt}}
\put(229,674.17){\rule{1.500pt}{0.400pt}}
\multiput(229.00,675.17)(3.887,-2.000){2}{\rule{0.750pt}{0.400pt}}
\multiput(236.00,672.95)(1.579,-0.447){3}{\rule{1.167pt}{0.108pt}}
\multiput(236.00,673.17)(5.579,-3.000){2}{\rule{0.583pt}{0.400pt}}
\multiput(244.00,669.95)(1.579,-0.447){3}{\rule{1.167pt}{0.108pt}}
\multiput(244.00,670.17)(5.579,-3.000){2}{\rule{0.583pt}{0.400pt}}
\put(252,666.17){\rule{1.700pt}{0.400pt}}
\multiput(252.00,667.17)(4.472,-2.000){2}{\rule{0.850pt}{0.400pt}}
\multiput(260.00,664.95)(1.579,-0.447){3}{\rule{1.167pt}{0.108pt}}
\multiput(260.00,665.17)(5.579,-3.000){2}{\rule{0.583pt}{0.400pt}}
\put(268,661.17){\rule{1.500pt}{0.400pt}}
\multiput(268.00,662.17)(3.887,-2.000){2}{\rule{0.750pt}{0.400pt}}
\multiput(275.00,659.95)(1.579,-0.447){3}{\rule{1.167pt}{0.108pt}}
\multiput(275.00,660.17)(5.579,-3.000){2}{\rule{0.583pt}{0.400pt}}
\put(283,656.17){\rule{1.700pt}{0.400pt}}
\multiput(283.00,657.17)(4.472,-2.000){2}{\rule{0.850pt}{0.400pt}}
\multiput(291.00,654.95)(1.579,-0.447){3}{\rule{1.167pt}{0.108pt}}
\multiput(291.00,655.17)(5.579,-3.000){2}{\rule{0.583pt}{0.400pt}}
\multiput(299.00,651.95)(1.355,-0.447){3}{\rule{1.033pt}{0.108pt}}
\multiput(299.00,652.17)(4.855,-3.000){2}{\rule{0.517pt}{0.400pt}}
\put(306,648.17){\rule{1.700pt}{0.400pt}}
\multiput(306.00,649.17)(4.472,-2.000){2}{\rule{0.850pt}{0.400pt}}
\multiput(314.00,646.95)(1.579,-0.447){3}{\rule{1.167pt}{0.108pt}}
\multiput(314.00,647.17)(5.579,-3.000){2}{\rule{0.583pt}{0.400pt}}
\put(322,643.17){\rule{1.700pt}{0.400pt}}
\multiput(322.00,644.17)(4.472,-2.000){2}{\rule{0.850pt}{0.400pt}}
\multiput(330.00,641.95)(1.579,-0.447){3}{\rule{1.167pt}{0.108pt}}
\multiput(330.00,642.17)(5.579,-3.000){2}{\rule{0.583pt}{0.400pt}}
\multiput(338.00,638.95)(1.355,-0.447){3}{\rule{1.033pt}{0.108pt}}
\multiput(338.00,639.17)(4.855,-3.000){2}{\rule{0.517pt}{0.400pt}}
\put(345,635.17){\rule{1.700pt}{0.400pt}}
\multiput(345.00,636.17)(4.472,-2.000){2}{\rule{0.850pt}{0.400pt}}
\multiput(353.00,633.95)(1.579,-0.447){3}{\rule{1.167pt}{0.108pt}}
\multiput(353.00,634.17)(5.579,-3.000){2}{\rule{0.583pt}{0.400pt}}
\put(361,630.17){\rule{1.700pt}{0.400pt}}
\multiput(361.00,631.17)(4.472,-2.000){2}{\rule{0.850pt}{0.400pt}}
\multiput(369.00,628.95)(1.355,-0.447){3}{\rule{1.033pt}{0.108pt}}
\multiput(369.00,629.17)(4.855,-3.000){2}{\rule{0.517pt}{0.400pt}}
\multiput(376.00,625.95)(1.579,-0.447){3}{\rule{1.167pt}{0.108pt}}
\multiput(376.00,626.17)(5.579,-3.000){2}{\rule{0.583pt}{0.400pt}}
\put(384,622.17){\rule{1.700pt}{0.400pt}}
\multiput(384.00,623.17)(4.472,-2.000){2}{\rule{0.850pt}{0.400pt}}
\multiput(392.00,620.95)(1.579,-0.447){3}{\rule{1.167pt}{0.108pt}}
\multiput(392.00,621.17)(5.579,-3.000){2}{\rule{0.583pt}{0.400pt}}
\put(400,617.17){\rule{1.700pt}{0.400pt}}
\multiput(400.00,618.17)(4.472,-2.000){2}{\rule{0.850pt}{0.400pt}}
\multiput(408.00,615.95)(1.355,-0.447){3}{\rule{1.033pt}{0.108pt}}
\multiput(408.00,616.17)(4.855,-3.000){2}{\rule{0.517pt}{0.400pt}}
\put(415,612.17){\rule{1.700pt}{0.400pt}}
\multiput(415.00,613.17)(4.472,-2.000){2}{\rule{0.850pt}{0.400pt}}
\multiput(423.00,610.95)(1.579,-0.447){3}{\rule{1.167pt}{0.108pt}}
\multiput(423.00,611.17)(5.579,-3.000){2}{\rule{0.583pt}{0.400pt}}
\multiput(431.00,607.95)(1.579,-0.447){3}{\rule{1.167pt}{0.108pt}}
\multiput(431.00,608.17)(5.579,-3.000){2}{\rule{0.583pt}{0.400pt}}
\put(439,604.17){\rule{1.500pt}{0.400pt}}
\multiput(439.00,605.17)(3.887,-2.000){2}{\rule{0.750pt}{0.400pt}}
\multiput(446.00,602.95)(1.579,-0.447){3}{\rule{1.167pt}{0.108pt}}
\multiput(446.00,603.17)(5.579,-3.000){2}{\rule{0.583pt}{0.400pt}}
\put(454,599.17){\rule{1.700pt}{0.400pt}}
\multiput(454.00,600.17)(4.472,-2.000){2}{\rule{0.850pt}{0.400pt}}
\multiput(462.00,597.95)(1.579,-0.447){3}{\rule{1.167pt}{0.108pt}}
\multiput(462.00,598.17)(5.579,-3.000){2}{\rule{0.583pt}{0.400pt}}
\multiput(470.00,594.95)(1.579,-0.447){3}{\rule{1.167pt}{0.108pt}}
\multiput(470.00,595.17)(5.579,-3.000){2}{\rule{0.583pt}{0.400pt}}
\put(478,591.17){\rule{1.500pt}{0.400pt}}
\multiput(478.00,592.17)(3.887,-2.000){2}{\rule{0.750pt}{0.400pt}}
\multiput(485.00,589.95)(1.579,-0.447){3}{\rule{1.167pt}{0.108pt}}
\multiput(485.00,590.17)(5.579,-3.000){2}{\rule{0.583pt}{0.400pt}}
\put(493,586.17){\rule{1.700pt}{0.400pt}}
\multiput(493.00,587.17)(4.472,-2.000){2}{\rule{0.850pt}{0.400pt}}
\multiput(501.00,584.95)(1.579,-0.447){3}{\rule{1.167pt}{0.108pt}}
\multiput(501.00,585.17)(5.579,-3.000){2}{\rule{0.583pt}{0.400pt}}
\put(509,581.17){\rule{1.500pt}{0.400pt}}
\multiput(509.00,582.17)(3.887,-2.000){2}{\rule{0.750pt}{0.400pt}}
\multiput(516.00,579.95)(1.579,-0.447){3}{\rule{1.167pt}{0.108pt}}
\multiput(516.00,580.17)(5.579,-3.000){2}{\rule{0.583pt}{0.400pt}}
\multiput(524.00,576.95)(1.579,-0.447){3}{\rule{1.167pt}{0.108pt}}
\multiput(524.00,577.17)(5.579,-3.000){2}{\rule{0.583pt}{0.400pt}}
\put(532,573.17){\rule{1.700pt}{0.400pt}}
\multiput(532.00,574.17)(4.472,-2.000){2}{\rule{0.850pt}{0.400pt}}
\multiput(540.00,571.95)(1.579,-0.447){3}{\rule{1.167pt}{0.108pt}}
\multiput(540.00,572.17)(5.579,-3.000){2}{\rule{0.583pt}{0.400pt}}
\put(548,568.17){\rule{1.500pt}{0.400pt}}
\multiput(548.00,569.17)(3.887,-2.000){2}{\rule{0.750pt}{0.400pt}}
\multiput(555.00,566.95)(1.579,-0.447){3}{\rule{1.167pt}{0.108pt}}
\multiput(555.00,567.17)(5.579,-3.000){2}{\rule{0.583pt}{0.400pt}}
\multiput(563.00,563.95)(1.579,-0.447){3}{\rule{1.167pt}{0.108pt}}
\multiput(563.00,564.17)(5.579,-3.000){2}{\rule{0.583pt}{0.400pt}}
\put(571,560.17){\rule{1.700pt}{0.400pt}}
\multiput(571.00,561.17)(4.472,-2.000){2}{\rule{0.850pt}{0.400pt}}
\multiput(579.00,558.95)(1.355,-0.447){3}{\rule{1.033pt}{0.108pt}}
\multiput(579.00,559.17)(4.855,-3.000){2}{\rule{0.517pt}{0.400pt}}
\put(586,555.17){\rule{1.700pt}{0.400pt}}
\multiput(586.00,556.17)(4.472,-2.000){2}{\rule{0.850pt}{0.400pt}}
\multiput(594.00,553.95)(1.579,-0.447){3}{\rule{1.167pt}{0.108pt}}
\multiput(594.00,554.17)(5.579,-3.000){2}{\rule{0.583pt}{0.400pt}}
\multiput(602.00,550.95)(1.579,-0.447){3}{\rule{1.167pt}{0.108pt}}
\multiput(602.00,551.17)(5.579,-3.000){2}{\rule{0.583pt}{0.400pt}}
\put(610,547.17){\rule{1.700pt}{0.400pt}}
\multiput(610.00,548.17)(4.472,-2.000){2}{\rule{0.850pt}{0.400pt}}
\multiput(618.00,545.95)(1.355,-0.447){3}{\rule{1.033pt}{0.108pt}}
\multiput(618.00,546.17)(4.855,-3.000){2}{\rule{0.517pt}{0.400pt}}
\put(625,542.17){\rule{1.700pt}{0.400pt}}
\multiput(625.00,543.17)(4.472,-2.000){2}{\rule{0.850pt}{0.400pt}}
\multiput(633.00,540.95)(1.579,-0.447){3}{\rule{1.167pt}{0.108pt}}
\multiput(633.00,541.17)(5.579,-3.000){2}{\rule{0.583pt}{0.400pt}}
\put(641,537.17){\rule{1.700pt}{0.400pt}}
\multiput(641.00,538.17)(4.472,-2.000){2}{\rule{0.850pt}{0.400pt}}
\multiput(649.00,535.95)(1.355,-0.447){3}{\rule{1.033pt}{0.108pt}}
\multiput(649.00,536.17)(4.855,-3.000){2}{\rule{0.517pt}{0.400pt}}
\multiput(656.00,532.95)(1.579,-0.447){3}{\rule{1.167pt}{0.108pt}}
\multiput(656.00,533.17)(5.579,-3.000){2}{\rule{0.583pt}{0.400pt}}
\put(664,529.17){\rule{1.700pt}{0.400pt}}
\multiput(664.00,530.17)(4.472,-2.000){2}{\rule{0.850pt}{0.400pt}}
\multiput(672.00,527.95)(1.579,-0.447){3}{\rule{1.167pt}{0.108pt}}
\multiput(672.00,528.17)(5.579,-3.000){2}{\rule{0.583pt}{0.400pt}}
\put(680,524.17){\rule{1.700pt}{0.400pt}}
\multiput(680.00,525.17)(4.472,-2.000){2}{\rule{0.850pt}{0.400pt}}
\multiput(688.00,522.95)(1.355,-0.447){3}{\rule{1.033pt}{0.108pt}}
\multiput(688.00,523.17)(4.855,-3.000){2}{\rule{0.517pt}{0.400pt}}
\multiput(695.00,519.95)(1.579,-0.447){3}{\rule{1.167pt}{0.108pt}}
\multiput(695.00,520.17)(5.579,-3.000){2}{\rule{0.583pt}{0.400pt}}
\put(703,516.17){\rule{1.700pt}{0.400pt}}
\multiput(703.00,517.17)(4.472,-2.000){2}{\rule{0.850pt}{0.400pt}}
\multiput(711.00,514.95)(1.579,-0.447){3}{\rule{1.167pt}{0.108pt}}
\multiput(711.00,515.17)(5.579,-3.000){2}{\rule{0.583pt}{0.400pt}}
\put(719,511.17){\rule{1.500pt}{0.400pt}}
\multiput(719.00,512.17)(3.887,-2.000){2}{\rule{0.750pt}{0.400pt}}
\multiput(726.00,509.95)(1.579,-0.447){3}{\rule{1.167pt}{0.108pt}}
\multiput(726.00,510.17)(5.579,-3.000){2}{\rule{0.583pt}{0.400pt}}
\put(734,506.17){\rule{1.700pt}{0.400pt}}
\multiput(734.00,507.17)(4.472,-2.000){2}{\rule{0.850pt}{0.400pt}}
\multiput(742.00,504.95)(1.579,-0.447){3}{\rule{1.167pt}{0.108pt}}
\multiput(742.00,505.17)(5.579,-3.000){2}{\rule{0.583pt}{0.400pt}}
\multiput(750.00,501.95)(1.579,-0.447){3}{\rule{1.167pt}{0.108pt}}
\multiput(750.00,502.17)(5.579,-3.000){2}{\rule{0.583pt}{0.400pt}}
\put(758,498.17){\rule{1.500pt}{0.400pt}}
\multiput(758.00,499.17)(3.887,-2.000){2}{\rule{0.750pt}{0.400pt}}
\multiput(765.00,496.95)(1.579,-0.447){3}{\rule{1.167pt}{0.108pt}}
\multiput(765.00,497.17)(5.579,-3.000){2}{\rule{0.583pt}{0.400pt}}
\put(773,493.17){\rule{1.700pt}{0.400pt}}
\multiput(773.00,494.17)(4.472,-2.000){2}{\rule{0.850pt}{0.400pt}}
\multiput(781.00,491.95)(1.579,-0.447){3}{\rule{1.167pt}{0.108pt}}
\multiput(781.00,492.17)(5.579,-3.000){2}{\rule{0.583pt}{0.400pt}}
\multiput(789.00,488.95)(1.355,-0.447){3}{\rule{1.033pt}{0.108pt}}
\multiput(789.00,489.17)(4.855,-3.000){2}{\rule{0.517pt}{0.400pt}}
\put(796,485.17){\rule{1.700pt}{0.400pt}}
\multiput(796.00,486.17)(4.472,-2.000){2}{\rule{0.850pt}{0.400pt}}
\multiput(804.00,483.95)(1.579,-0.447){3}{\rule{1.167pt}{0.108pt}}
\multiput(804.00,484.17)(5.579,-3.000){2}{\rule{0.583pt}{0.400pt}}
\put(812,480.17){\rule{1.700pt}{0.400pt}}
\multiput(812.00,481.17)(4.472,-2.000){2}{\rule{0.850pt}{0.400pt}}
\multiput(820.00,478.95)(1.579,-0.447){3}{\rule{1.167pt}{0.108pt}}
\multiput(820.00,479.17)(5.579,-3.000){2}{\rule{0.583pt}{0.400pt}}
\multiput(828.00,475.95)(1.355,-0.447){3}{\rule{1.033pt}{0.108pt}}
\multiput(828.00,476.17)(4.855,-3.000){2}{\rule{0.517pt}{0.400pt}}
\put(835,472.17){\rule{1.700pt}{0.400pt}}
\multiput(835.00,473.17)(4.472,-2.000){2}{\rule{0.850pt}{0.400pt}}
\multiput(843.00,470.95)(1.579,-0.447){3}{\rule{1.167pt}{0.108pt}}
\multiput(843.00,471.17)(5.579,-3.000){2}{\rule{0.583pt}{0.400pt}}
\put(851,467.17){\rule{1.700pt}{0.400pt}}
\multiput(851.00,468.17)(4.472,-2.000){2}{\rule{0.850pt}{0.400pt}}
\multiput(859.00,465.95)(1.355,-0.447){3}{\rule{1.033pt}{0.108pt}}
\multiput(859.00,466.17)(4.855,-3.000){2}{\rule{0.517pt}{0.400pt}}
\put(866,462.17){\rule{1.700pt}{0.400pt}}
\multiput(866.00,463.17)(4.472,-2.000){2}{\rule{0.850pt}{0.400pt}}
\multiput(874.00,460.95)(1.579,-0.447){3}{\rule{1.167pt}{0.108pt}}
\multiput(874.00,461.17)(5.579,-3.000){2}{\rule{0.583pt}{0.400pt}}
\multiput(882.00,457.95)(1.579,-0.447){3}{\rule{1.167pt}{0.108pt}}
\multiput(882.00,458.17)(5.579,-3.000){2}{\rule{0.583pt}{0.400pt}}
\put(890,454.17){\rule{1.700pt}{0.400pt}}
\multiput(890.00,455.17)(4.472,-2.000){2}{\rule{0.850pt}{0.400pt}}
\multiput(898.00,452.95)(1.355,-0.447){3}{\rule{1.033pt}{0.108pt}}
\multiput(898.00,453.17)(4.855,-3.000){2}{\rule{0.517pt}{0.400pt}}
\put(905,449.17){\rule{1.700pt}{0.400pt}}
\multiput(905.00,450.17)(4.472,-2.000){2}{\rule{0.850pt}{0.400pt}}
\multiput(913.00,447.95)(1.579,-0.447){3}{\rule{1.167pt}{0.108pt}}
\multiput(913.00,448.17)(5.579,-3.000){2}{\rule{0.583pt}{0.400pt}}
\multiput(921.00,444.95)(1.579,-0.447){3}{\rule{1.167pt}{0.108pt}}
\multiput(921.00,445.17)(5.579,-3.000){2}{\rule{0.583pt}{0.400pt}}
\put(929,441.17){\rule{1.500pt}{0.400pt}}
\multiput(929.00,442.17)(3.887,-2.000){2}{\rule{0.750pt}{0.400pt}}
\multiput(936.00,439.95)(1.579,-0.447){3}{\rule{1.167pt}{0.108pt}}
\multiput(936.00,440.17)(5.579,-3.000){2}{\rule{0.583pt}{0.400pt}}
\put(944,436.17){\rule{1.700pt}{0.400pt}}
\multiput(944.00,437.17)(4.472,-2.000){2}{\rule{0.850pt}{0.400pt}}
}{\color{green}\put(182,439){\usebox{\plotpoint}}
\put(182,438.67){\rule{1.927pt}{0.400pt}}
\multiput(182.00,438.17)(4.000,1.000){2}{\rule{0.964pt}{0.400pt}}
\put(198,439.67){\rule{1.686pt}{0.400pt}}
\multiput(198.00,439.17)(3.500,1.000){2}{\rule{0.843pt}{0.400pt}}
\put(190.0,440.0){\rule[-0.200pt]{1.927pt}{0.400pt}}
\put(213,440.67){\rule{1.927pt}{0.400pt}}
\multiput(213.00,440.17)(4.000,1.000){2}{\rule{0.964pt}{0.400pt}}
\put(221,441.67){\rule{1.927pt}{0.400pt}}
\multiput(221.00,441.17)(4.000,1.000){2}{\rule{0.964pt}{0.400pt}}
\put(205.0,441.0){\rule[-0.200pt]{1.927pt}{0.400pt}}
\put(236,442.67){\rule{1.927pt}{0.400pt}}
\multiput(236.00,442.17)(4.000,1.000){2}{\rule{0.964pt}{0.400pt}}
\put(229.0,443.0){\rule[-0.200pt]{1.686pt}{0.400pt}}
\put(252,443.67){\rule{1.927pt}{0.400pt}}
\multiput(252.00,443.17)(4.000,1.000){2}{\rule{0.964pt}{0.400pt}}
\put(244.0,444.0){\rule[-0.200pt]{1.927pt}{0.400pt}}
\put(268,444.67){\rule{1.686pt}{0.400pt}}
\multiput(268.00,444.17)(3.500,1.000){2}{\rule{0.843pt}{0.400pt}}
\put(275,445.67){\rule{1.927pt}{0.400pt}}
\multiput(275.00,445.17)(4.000,1.000){2}{\rule{0.964pt}{0.400pt}}
\put(260.0,445.0){\rule[-0.200pt]{1.927pt}{0.400pt}}
\put(291,446.67){\rule{1.927pt}{0.400pt}}
\multiput(291.00,446.17)(4.000,1.000){2}{\rule{0.964pt}{0.400pt}}
\put(283.0,447.0){\rule[-0.200pt]{1.927pt}{0.400pt}}
\put(306,447.67){\rule{1.927pt}{0.400pt}}
\multiput(306.00,447.17)(4.000,1.000){2}{\rule{0.964pt}{0.400pt}}
\put(299.0,448.0){\rule[-0.200pt]{1.686pt}{0.400pt}}
\put(322,448.67){\rule{1.927pt}{0.400pt}}
\multiput(322.00,448.17)(4.000,1.000){2}{\rule{0.964pt}{0.400pt}}
\put(330,449.67){\rule{1.927pt}{0.400pt}}
\multiput(330.00,449.17)(4.000,1.000){2}{\rule{0.964pt}{0.400pt}}
\put(314.0,449.0){\rule[-0.200pt]{1.927pt}{0.400pt}}
\put(345,450.67){\rule{1.927pt}{0.400pt}}
\multiput(345.00,450.17)(4.000,1.000){2}{\rule{0.964pt}{0.400pt}}
\put(338.0,451.0){\rule[-0.200pt]{1.686pt}{0.400pt}}
\put(361,451.67){\rule{1.927pt}{0.400pt}}
\multiput(361.00,451.17)(4.000,1.000){2}{\rule{0.964pt}{0.400pt}}
\put(353.0,452.0){\rule[-0.200pt]{1.927pt}{0.400pt}}
\put(376,452.67){\rule{1.927pt}{0.400pt}}
\multiput(376.00,452.17)(4.000,1.000){2}{\rule{0.964pt}{0.400pt}}
\put(384,453.67){\rule{1.927pt}{0.400pt}}
\multiput(384.00,453.17)(4.000,1.000){2}{\rule{0.964pt}{0.400pt}}
\put(369.0,453.0){\rule[-0.200pt]{1.686pt}{0.400pt}}
\put(400,454.67){\rule{1.927pt}{0.400pt}}
\multiput(400.00,454.17)(4.000,1.000){2}{\rule{0.964pt}{0.400pt}}
\put(392.0,455.0){\rule[-0.200pt]{1.927pt}{0.400pt}}
\put(415,455.67){\rule{1.927pt}{0.400pt}}
\multiput(415.00,455.17)(4.000,1.000){2}{\rule{0.964pt}{0.400pt}}
\put(408.0,456.0){\rule[-0.200pt]{1.686pt}{0.400pt}}
\put(431,456.67){\rule{1.927pt}{0.400pt}}
\multiput(431.00,456.17)(4.000,1.000){2}{\rule{0.964pt}{0.400pt}}
\put(439,457.67){\rule{1.686pt}{0.400pt}}
\multiput(439.00,457.17)(3.500,1.000){2}{\rule{0.843pt}{0.400pt}}
\put(423.0,457.0){\rule[-0.200pt]{1.927pt}{0.400pt}}
\put(454,458.67){\rule{1.927pt}{0.400pt}}
\multiput(454.00,458.17)(4.000,1.000){2}{\rule{0.964pt}{0.400pt}}
\put(446.0,459.0){\rule[-0.200pt]{1.927pt}{0.400pt}}
\put(470,459.67){\rule{1.927pt}{0.400pt}}
\multiput(470.00,459.17)(4.000,1.000){2}{\rule{0.964pt}{0.400pt}}
\put(462.0,460.0){\rule[-0.200pt]{1.927pt}{0.400pt}}
\put(485,460.67){\rule{1.927pt}{0.400pt}}
\multiput(485.00,460.17)(4.000,1.000){2}{\rule{0.964pt}{0.400pt}}
\put(493,461.67){\rule{1.927pt}{0.400pt}}
\multiput(493.00,461.17)(4.000,1.000){2}{\rule{0.964pt}{0.400pt}}
\put(478.0,461.0){\rule[-0.200pt]{1.686pt}{0.400pt}}
\put(509,462.67){\rule{1.686pt}{0.400pt}}
\multiput(509.00,462.17)(3.500,1.000){2}{\rule{0.843pt}{0.400pt}}
\put(501.0,463.0){\rule[-0.200pt]{1.927pt}{0.400pt}}
\put(524,463.67){\rule{1.927pt}{0.400pt}}
\multiput(524.00,463.17)(4.000,1.000){2}{\rule{0.964pt}{0.400pt}}
\put(532,464.67){\rule{1.927pt}{0.400pt}}
\multiput(532.00,464.17)(4.000,1.000){2}{\rule{0.964pt}{0.400pt}}
\put(516.0,464.0){\rule[-0.200pt]{1.927pt}{0.400pt}}
\put(548,465.67){\rule{1.686pt}{0.400pt}}
\multiput(548.00,465.17)(3.500,1.000){2}{\rule{0.843pt}{0.400pt}}
\put(540.0,466.0){\rule[-0.200pt]{1.927pt}{0.400pt}}
\put(563,466.67){\rule{1.927pt}{0.400pt}}
\multiput(563.00,466.17)(4.000,1.000){2}{\rule{0.964pt}{0.400pt}}
\put(555.0,467.0){\rule[-0.200pt]{1.927pt}{0.400pt}}
\put(579,467.67){\rule{1.686pt}{0.400pt}}
\multiput(579.00,467.17)(3.500,1.000){2}{\rule{0.843pt}{0.400pt}}
\put(586,468.67){\rule{1.927pt}{0.400pt}}
\multiput(586.00,468.17)(4.000,1.000){2}{\rule{0.964pt}{0.400pt}}
\put(571.0,468.0){\rule[-0.200pt]{1.927pt}{0.400pt}}
\put(602,469.67){\rule{1.927pt}{0.400pt}}
\multiput(602.00,469.17)(4.000,1.000){2}{\rule{0.964pt}{0.400pt}}
\put(594.0,470.0){\rule[-0.200pt]{1.927pt}{0.400pt}}
\put(618,470.67){\rule{1.686pt}{0.400pt}}
\multiput(618.00,470.17)(3.500,1.000){2}{\rule{0.843pt}{0.400pt}}
\put(610.0,471.0){\rule[-0.200pt]{1.927pt}{0.400pt}}
\put(633,471.67){\rule{1.927pt}{0.400pt}}
\multiput(633.00,471.17)(4.000,1.000){2}{\rule{0.964pt}{0.400pt}}
\put(641,472.67){\rule{1.927pt}{0.400pt}}
\multiput(641.00,472.17)(4.000,1.000){2}{\rule{0.964pt}{0.400pt}}
\put(625.0,472.0){\rule[-0.200pt]{1.927pt}{0.400pt}}
\put(656,473.67){\rule{1.927pt}{0.400pt}}
\multiput(656.00,473.17)(4.000,1.000){2}{\rule{0.964pt}{0.400pt}}
\put(649.0,474.0){\rule[-0.200pt]{1.686pt}{0.400pt}}
\put(672,474.67){\rule{1.927pt}{0.400pt}}
\multiput(672.00,474.17)(4.000,1.000){2}{\rule{0.964pt}{0.400pt}}
\put(664.0,475.0){\rule[-0.200pt]{1.927pt}{0.400pt}}
\put(688,475.67){\rule{1.686pt}{0.400pt}}
\multiput(688.00,475.17)(3.500,1.000){2}{\rule{0.843pt}{0.400pt}}
\put(695,476.67){\rule{1.927pt}{0.400pt}}
\multiput(695.00,476.17)(4.000,1.000){2}{\rule{0.964pt}{0.400pt}}
\put(680.0,476.0){\rule[-0.200pt]{1.927pt}{0.400pt}}
\put(711,477.67){\rule{1.927pt}{0.400pt}}
\multiput(711.00,477.17)(4.000,1.000){2}{\rule{0.964pt}{0.400pt}}
\put(703.0,478.0){\rule[-0.200pt]{1.927pt}{0.400pt}}
\put(726,478.67){\rule{1.927pt}{0.400pt}}
\multiput(726.00,478.17)(4.000,1.000){2}{\rule{0.964pt}{0.400pt}}
\put(719.0,479.0){\rule[-0.200pt]{1.686pt}{0.400pt}}
\put(742,479.67){\rule{1.927pt}{0.400pt}}
\multiput(742.00,479.17)(4.000,1.000){2}{\rule{0.964pt}{0.400pt}}
\put(750,480.67){\rule{1.927pt}{0.400pt}}
\multiput(750.00,480.17)(4.000,1.000){2}{\rule{0.964pt}{0.400pt}}
\put(734.0,480.0){\rule[-0.200pt]{1.927pt}{0.400pt}}
\put(765,481.67){\rule{1.927pt}{0.400pt}}
\multiput(765.00,481.17)(4.000,1.000){2}{\rule{0.964pt}{0.400pt}}
\put(758.0,482.0){\rule[-0.200pt]{1.686pt}{0.400pt}}
\put(781,482.67){\rule{1.927pt}{0.400pt}}
\multiput(781.00,482.17)(4.000,1.000){2}{\rule{0.964pt}{0.400pt}}
\put(773.0,483.0){\rule[-0.200pt]{1.927pt}{0.400pt}}
\put(796,483.67){\rule{1.927pt}{0.400pt}}
\multiput(796.00,483.17)(4.000,1.000){2}{\rule{0.964pt}{0.400pt}}
\put(804,484.67){\rule{1.927pt}{0.400pt}}
\multiput(804.00,484.17)(4.000,1.000){2}{\rule{0.964pt}{0.400pt}}
\put(789.0,484.0){\rule[-0.200pt]{1.686pt}{0.400pt}}
\put(820,485.67){\rule{1.927pt}{0.400pt}}
\multiput(820.00,485.17)(4.000,1.000){2}{\rule{0.964pt}{0.400pt}}
\put(812.0,486.0){\rule[-0.200pt]{1.927pt}{0.400pt}}
\put(835,486.67){\rule{1.927pt}{0.400pt}}
\multiput(835.00,486.17)(4.000,1.000){2}{\rule{0.964pt}{0.400pt}}
\put(843,487.67){\rule{1.927pt}{0.400pt}}
\multiput(843.00,487.17)(4.000,1.000){2}{\rule{0.964pt}{0.400pt}}
\put(828.0,487.0){\rule[-0.200pt]{1.686pt}{0.400pt}}
\put(859,488.67){\rule{1.686pt}{0.400pt}}
\multiput(859.00,488.17)(3.500,1.000){2}{\rule{0.843pt}{0.400pt}}
\put(851.0,489.0){\rule[-0.200pt]{1.927pt}{0.400pt}}
\put(874,489.67){\rule{1.927pt}{0.400pt}}
\multiput(874.00,489.17)(4.000,1.000){2}{\rule{0.964pt}{0.400pt}}
\put(866.0,490.0){\rule[-0.200pt]{1.927pt}{0.400pt}}
\put(890,490.67){\rule{1.927pt}{0.400pt}}
\multiput(890.00,490.17)(4.000,1.000){2}{\rule{0.964pt}{0.400pt}}
\put(898,491.67){\rule{1.686pt}{0.400pt}}
\multiput(898.00,491.17)(3.500,1.000){2}{\rule{0.843pt}{0.400pt}}
\put(882.0,491.0){\rule[-0.200pt]{1.927pt}{0.400pt}}
\put(913,492.67){\rule{1.927pt}{0.400pt}}
\multiput(913.00,492.17)(4.000,1.000){2}{\rule{0.964pt}{0.400pt}}
\put(905.0,493.0){\rule[-0.200pt]{1.927pt}{0.400pt}}
\put(929,493.67){\rule{1.686pt}{0.400pt}}
\multiput(929.00,493.17)(3.500,1.000){2}{\rule{0.843pt}{0.400pt}}
\put(921.0,494.0){\rule[-0.200pt]{1.927pt}{0.400pt}}
\put(944,494.67){\rule{1.927pt}{0.400pt}}
\multiput(944.00,494.17)(4.000,1.000){2}{\rule{0.964pt}{0.400pt}}
\put(936.0,495.0){\rule[-0.200pt]{1.927pt}{0.400pt}}
}{\color{red}\put(182,170){\usebox{\plotpoint}}
\multiput(182.00,170.60)(1.066,0.468){5}{\rule{0.900pt}{0.113pt}}
\multiput(182.00,169.17)(6.132,4.000){2}{\rule{0.450pt}{0.400pt}}
\multiput(190.00,174.60)(1.066,0.468){5}{\rule{0.900pt}{0.113pt}}
\multiput(190.00,173.17)(6.132,4.000){2}{\rule{0.450pt}{0.400pt}}
\multiput(198.00,178.60)(0.920,0.468){5}{\rule{0.800pt}{0.113pt}}
\multiput(198.00,177.17)(5.340,4.000){2}{\rule{0.400pt}{0.400pt}}
\multiput(205.00,182.60)(1.066,0.468){5}{\rule{0.900pt}{0.113pt}}
\multiput(205.00,181.17)(6.132,4.000){2}{\rule{0.450pt}{0.400pt}}
\multiput(213.00,186.60)(1.066,0.468){5}{\rule{0.900pt}{0.113pt}}
\multiput(213.00,185.17)(6.132,4.000){2}{\rule{0.450pt}{0.400pt}}
\multiput(221.00,190.60)(1.066,0.468){5}{\rule{0.900pt}{0.113pt}}
\multiput(221.00,189.17)(6.132,4.000){2}{\rule{0.450pt}{0.400pt}}
\multiput(229.00,194.60)(0.920,0.468){5}{\rule{0.800pt}{0.113pt}}
\multiput(229.00,193.17)(5.340,4.000){2}{\rule{0.400pt}{0.400pt}}
\multiput(236.00,198.60)(1.066,0.468){5}{\rule{0.900pt}{0.113pt}}
\multiput(236.00,197.17)(6.132,4.000){2}{\rule{0.450pt}{0.400pt}}
\multiput(244.00,202.60)(1.066,0.468){5}{\rule{0.900pt}{0.113pt}}
\multiput(244.00,201.17)(6.132,4.000){2}{\rule{0.450pt}{0.400pt}}
\multiput(252.00,206.60)(1.066,0.468){5}{\rule{0.900pt}{0.113pt}}
\multiput(252.00,205.17)(6.132,4.000){2}{\rule{0.450pt}{0.400pt}}
\multiput(260.00,210.59)(0.821,0.477){7}{\rule{0.740pt}{0.115pt}}
\multiput(260.00,209.17)(6.464,5.000){2}{\rule{0.370pt}{0.400pt}}
\multiput(268.00,215.60)(0.920,0.468){5}{\rule{0.800pt}{0.113pt}}
\multiput(268.00,214.17)(5.340,4.000){2}{\rule{0.400pt}{0.400pt}}
\multiput(275.00,219.60)(1.066,0.468){5}{\rule{0.900pt}{0.113pt}}
\multiput(275.00,218.17)(6.132,4.000){2}{\rule{0.450pt}{0.400pt}}
\multiput(283.00,223.60)(1.066,0.468){5}{\rule{0.900pt}{0.113pt}}
\multiput(283.00,222.17)(6.132,4.000){2}{\rule{0.450pt}{0.400pt}}
\multiput(291.00,227.60)(1.066,0.468){5}{\rule{0.900pt}{0.113pt}}
\multiput(291.00,226.17)(6.132,4.000){2}{\rule{0.450pt}{0.400pt}}
\multiput(299.00,231.60)(0.920,0.468){5}{\rule{0.800pt}{0.113pt}}
\multiput(299.00,230.17)(5.340,4.000){2}{\rule{0.400pt}{0.400pt}}
\multiput(306.00,235.60)(1.066,0.468){5}{\rule{0.900pt}{0.113pt}}
\multiput(306.00,234.17)(6.132,4.000){2}{\rule{0.450pt}{0.400pt}}
\multiput(314.00,239.60)(1.066,0.468){5}{\rule{0.900pt}{0.113pt}}
\multiput(314.00,238.17)(6.132,4.000){2}{\rule{0.450pt}{0.400pt}}
\multiput(322.00,243.60)(1.066,0.468){5}{\rule{0.900pt}{0.113pt}}
\multiput(322.00,242.17)(6.132,4.000){2}{\rule{0.450pt}{0.400pt}}
\multiput(330.00,247.60)(1.066,0.468){5}{\rule{0.900pt}{0.113pt}}
\multiput(330.00,246.17)(6.132,4.000){2}{\rule{0.450pt}{0.400pt}}
\multiput(338.00,251.60)(0.920,0.468){5}{\rule{0.800pt}{0.113pt}}
\multiput(338.00,250.17)(5.340,4.000){2}{\rule{0.400pt}{0.400pt}}
\multiput(345.00,255.60)(1.066,0.468){5}{\rule{0.900pt}{0.113pt}}
\multiput(345.00,254.17)(6.132,4.000){2}{\rule{0.450pt}{0.400pt}}
\multiput(353.00,259.60)(1.066,0.468){5}{\rule{0.900pt}{0.113pt}}
\multiput(353.00,258.17)(6.132,4.000){2}{\rule{0.450pt}{0.400pt}}
\multiput(361.00,263.60)(1.066,0.468){5}{\rule{0.900pt}{0.113pt}}
\multiput(361.00,262.17)(6.132,4.000){2}{\rule{0.450pt}{0.400pt}}
\multiput(369.00,267.60)(0.920,0.468){5}{\rule{0.800pt}{0.113pt}}
\multiput(369.00,266.17)(5.340,4.000){2}{\rule{0.400pt}{0.400pt}}
\multiput(376.00,271.60)(1.066,0.468){5}{\rule{0.900pt}{0.113pt}}
\multiput(376.00,270.17)(6.132,4.000){2}{\rule{0.450pt}{0.400pt}}
\multiput(384.00,275.60)(1.066,0.468){5}{\rule{0.900pt}{0.113pt}}
\multiput(384.00,274.17)(6.132,4.000){2}{\rule{0.450pt}{0.400pt}}
\multiput(392.00,279.60)(1.066,0.468){5}{\rule{0.900pt}{0.113pt}}
\multiput(392.00,278.17)(6.132,4.000){2}{\rule{0.450pt}{0.400pt}}
\multiput(400.00,283.60)(1.066,0.468){5}{\rule{0.900pt}{0.113pt}}
\multiput(400.00,282.17)(6.132,4.000){2}{\rule{0.450pt}{0.400pt}}
\multiput(408.00,287.60)(0.920,0.468){5}{\rule{0.800pt}{0.113pt}}
\multiput(408.00,286.17)(5.340,4.000){2}{\rule{0.400pt}{0.400pt}}
\multiput(415.00,291.60)(1.066,0.468){5}{\rule{0.900pt}{0.113pt}}
\multiput(415.00,290.17)(6.132,4.000){2}{\rule{0.450pt}{0.400pt}}
\multiput(423.00,295.60)(1.066,0.468){5}{\rule{0.900pt}{0.113pt}}
\multiput(423.00,294.17)(6.132,4.000){2}{\rule{0.450pt}{0.400pt}}
\multiput(431.00,299.60)(1.066,0.468){5}{\rule{0.900pt}{0.113pt}}
\multiput(431.00,298.17)(6.132,4.000){2}{\rule{0.450pt}{0.400pt}}
\multiput(439.00,303.60)(0.920,0.468){5}{\rule{0.800pt}{0.113pt}}
\multiput(439.00,302.17)(5.340,4.000){2}{\rule{0.400pt}{0.400pt}}
\multiput(446.00,307.60)(1.066,0.468){5}{\rule{0.900pt}{0.113pt}}
\multiput(446.00,306.17)(6.132,4.000){2}{\rule{0.450pt}{0.400pt}}
\multiput(454.00,311.60)(1.066,0.468){5}{\rule{0.900pt}{0.113pt}}
\multiput(454.00,310.17)(6.132,4.000){2}{\rule{0.450pt}{0.400pt}}
\multiput(462.00,315.60)(1.066,0.468){5}{\rule{0.900pt}{0.113pt}}
\multiput(462.00,314.17)(6.132,4.000){2}{\rule{0.450pt}{0.400pt}}
\multiput(470.00,319.60)(1.066,0.468){5}{\rule{0.900pt}{0.113pt}}
\multiput(470.00,318.17)(6.132,4.000){2}{\rule{0.450pt}{0.400pt}}
\multiput(478.00,323.60)(0.920,0.468){5}{\rule{0.800pt}{0.113pt}}
\multiput(478.00,322.17)(5.340,4.000){2}{\rule{0.400pt}{0.400pt}}
\multiput(485.00,327.60)(1.066,0.468){5}{\rule{0.900pt}{0.113pt}}
\multiput(485.00,326.17)(6.132,4.000){2}{\rule{0.450pt}{0.400pt}}
\multiput(493.00,331.60)(1.066,0.468){5}{\rule{0.900pt}{0.113pt}}
\multiput(493.00,330.17)(6.132,4.000){2}{\rule{0.450pt}{0.400pt}}
\multiput(501.00,335.60)(1.066,0.468){5}{\rule{0.900pt}{0.113pt}}
\multiput(501.00,334.17)(6.132,4.000){2}{\rule{0.450pt}{0.400pt}}
\multiput(509.00,339.60)(0.920,0.468){5}{\rule{0.800pt}{0.113pt}}
\multiput(509.00,338.17)(5.340,4.000){2}{\rule{0.400pt}{0.400pt}}
\multiput(516.00,343.60)(1.066,0.468){5}{\rule{0.900pt}{0.113pt}}
\multiput(516.00,342.17)(6.132,4.000){2}{\rule{0.450pt}{0.400pt}}
\multiput(524.00,347.60)(1.066,0.468){5}{\rule{0.900pt}{0.113pt}}
\multiput(524.00,346.17)(6.132,4.000){2}{\rule{0.450pt}{0.400pt}}
\multiput(532.00,351.60)(1.066,0.468){5}{\rule{0.900pt}{0.113pt}}
\multiput(532.00,350.17)(6.132,4.000){2}{\rule{0.450pt}{0.400pt}}
\multiput(540.00,355.60)(1.066,0.468){5}{\rule{0.900pt}{0.113pt}}
\multiput(540.00,354.17)(6.132,4.000){2}{\rule{0.450pt}{0.400pt}}
\multiput(548.00,359.60)(0.920,0.468){5}{\rule{0.800pt}{0.113pt}}
\multiput(548.00,358.17)(5.340,4.000){2}{\rule{0.400pt}{0.400pt}}
\multiput(555.00,363.60)(1.066,0.468){5}{\rule{0.900pt}{0.113pt}}
\multiput(555.00,362.17)(6.132,4.000){2}{\rule{0.450pt}{0.400pt}}
\multiput(563.00,367.60)(1.066,0.468){5}{\rule{0.900pt}{0.113pt}}
\multiput(563.00,366.17)(6.132,4.000){2}{\rule{0.450pt}{0.400pt}}
\multiput(571.00,371.60)(1.066,0.468){5}{\rule{0.900pt}{0.113pt}}
\multiput(571.00,370.17)(6.132,4.000){2}{\rule{0.450pt}{0.400pt}}
\multiput(579.00,375.60)(0.920,0.468){5}{\rule{0.800pt}{0.113pt}}
\multiput(579.00,374.17)(5.340,4.000){2}{\rule{0.400pt}{0.400pt}}
\multiput(586.00,379.59)(0.821,0.477){7}{\rule{0.740pt}{0.115pt}}
\multiput(586.00,378.17)(6.464,5.000){2}{\rule{0.370pt}{0.400pt}}
\multiput(594.00,384.60)(1.066,0.468){5}{\rule{0.900pt}{0.113pt}}
\multiput(594.00,383.17)(6.132,4.000){2}{\rule{0.450pt}{0.400pt}}
\multiput(602.00,388.60)(1.066,0.468){5}{\rule{0.900pt}{0.113pt}}
\multiput(602.00,387.17)(6.132,4.000){2}{\rule{0.450pt}{0.400pt}}
\multiput(610.00,392.60)(1.066,0.468){5}{\rule{0.900pt}{0.113pt}}
\multiput(610.00,391.17)(6.132,4.000){2}{\rule{0.450pt}{0.400pt}}
\multiput(618.00,396.60)(0.920,0.468){5}{\rule{0.800pt}{0.113pt}}
\multiput(618.00,395.17)(5.340,4.000){2}{\rule{0.400pt}{0.400pt}}
\multiput(625.00,400.60)(1.066,0.468){5}{\rule{0.900pt}{0.113pt}}
\multiput(625.00,399.17)(6.132,4.000){2}{\rule{0.450pt}{0.400pt}}
\multiput(633.00,404.60)(1.066,0.468){5}{\rule{0.900pt}{0.113pt}}
\multiput(633.00,403.17)(6.132,4.000){2}{\rule{0.450pt}{0.400pt}}
\multiput(641.00,408.60)(1.066,0.468){5}{\rule{0.900pt}{0.113pt}}
\multiput(641.00,407.17)(6.132,4.000){2}{\rule{0.450pt}{0.400pt}}
\multiput(649.00,412.60)(0.920,0.468){5}{\rule{0.800pt}{0.113pt}}
\multiput(649.00,411.17)(5.340,4.000){2}{\rule{0.400pt}{0.400pt}}
\multiput(656.00,416.60)(1.066,0.468){5}{\rule{0.900pt}{0.113pt}}
\multiput(656.00,415.17)(6.132,4.000){2}{\rule{0.450pt}{0.400pt}}
\multiput(664.00,420.60)(1.066,0.468){5}{\rule{0.900pt}{0.113pt}}
\multiput(664.00,419.17)(6.132,4.000){2}{\rule{0.450pt}{0.400pt}}
\multiput(672.00,424.60)(1.066,0.468){5}{\rule{0.900pt}{0.113pt}}
\multiput(672.00,423.17)(6.132,4.000){2}{\rule{0.450pt}{0.400pt}}
\multiput(680.00,428.60)(1.066,0.468){5}{\rule{0.900pt}{0.113pt}}
\multiput(680.00,427.17)(6.132,4.000){2}{\rule{0.450pt}{0.400pt}}
\multiput(688.00,432.60)(0.920,0.468){5}{\rule{0.800pt}{0.113pt}}
\multiput(688.00,431.17)(5.340,4.000){2}{\rule{0.400pt}{0.400pt}}
\multiput(695.00,436.60)(1.066,0.468){5}{\rule{0.900pt}{0.113pt}}
\multiput(695.00,435.17)(6.132,4.000){2}{\rule{0.450pt}{0.400pt}}
\multiput(703.00,440.60)(1.066,0.468){5}{\rule{0.900pt}{0.113pt}}
\multiput(703.00,439.17)(6.132,4.000){2}{\rule{0.450pt}{0.400pt}}
\multiput(711.00,444.60)(1.066,0.468){5}{\rule{0.900pt}{0.113pt}}
\multiput(711.00,443.17)(6.132,4.000){2}{\rule{0.450pt}{0.400pt}}
\multiput(719.00,448.60)(0.920,0.468){5}{\rule{0.800pt}{0.113pt}}
\multiput(719.00,447.17)(5.340,4.000){2}{\rule{0.400pt}{0.400pt}}
\multiput(726.00,452.60)(1.066,0.468){5}{\rule{0.900pt}{0.113pt}}
\multiput(726.00,451.17)(6.132,4.000){2}{\rule{0.450pt}{0.400pt}}
\multiput(734.00,456.60)(1.066,0.468){5}{\rule{0.900pt}{0.113pt}}
\multiput(734.00,455.17)(6.132,4.000){2}{\rule{0.450pt}{0.400pt}}
\multiput(742.00,460.60)(1.066,0.468){5}{\rule{0.900pt}{0.113pt}}
\multiput(742.00,459.17)(6.132,4.000){2}{\rule{0.450pt}{0.400pt}}
\multiput(750.00,464.60)(1.066,0.468){5}{\rule{0.900pt}{0.113pt}}
\multiput(750.00,463.17)(6.132,4.000){2}{\rule{0.450pt}{0.400pt}}
\multiput(758.00,468.60)(0.920,0.468){5}{\rule{0.800pt}{0.113pt}}
\multiput(758.00,467.17)(5.340,4.000){2}{\rule{0.400pt}{0.400pt}}
\multiput(765.00,472.60)(1.066,0.468){5}{\rule{0.900pt}{0.113pt}}
\multiput(765.00,471.17)(6.132,4.000){2}{\rule{0.450pt}{0.400pt}}
\multiput(773.00,476.60)(1.066,0.468){5}{\rule{0.900pt}{0.113pt}}
\multiput(773.00,475.17)(6.132,4.000){2}{\rule{0.450pt}{0.400pt}}
\multiput(781.00,480.60)(1.066,0.468){5}{\rule{0.900pt}{0.113pt}}
\multiput(781.00,479.17)(6.132,4.000){2}{\rule{0.450pt}{0.400pt}}
\multiput(789.00,484.60)(0.920,0.468){5}{\rule{0.800pt}{0.113pt}}
\multiput(789.00,483.17)(5.340,4.000){2}{\rule{0.400pt}{0.400pt}}
\multiput(796.00,488.60)(1.066,0.468){5}{\rule{0.900pt}{0.113pt}}
\multiput(796.00,487.17)(6.132,4.000){2}{\rule{0.450pt}{0.400pt}}
\multiput(804.00,492.60)(1.066,0.468){5}{\rule{0.900pt}{0.113pt}}
\multiput(804.00,491.17)(6.132,4.000){2}{\rule{0.450pt}{0.400pt}}
\multiput(812.00,496.60)(1.066,0.468){5}{\rule{0.900pt}{0.113pt}}
\multiput(812.00,495.17)(6.132,4.000){2}{\rule{0.450pt}{0.400pt}}
\multiput(820.00,500.60)(1.066,0.468){5}{\rule{0.900pt}{0.113pt}}
\multiput(820.00,499.17)(6.132,4.000){2}{\rule{0.450pt}{0.400pt}}
\multiput(828.00,504.60)(0.920,0.468){5}{\rule{0.800pt}{0.113pt}}
\multiput(828.00,503.17)(5.340,4.000){2}{\rule{0.400pt}{0.400pt}}
\multiput(835.00,508.60)(1.066,0.468){5}{\rule{0.900pt}{0.113pt}}
\multiput(835.00,507.17)(6.132,4.000){2}{\rule{0.450pt}{0.400pt}}
\multiput(843.00,512.60)(1.066,0.468){5}{\rule{0.900pt}{0.113pt}}
\multiput(843.00,511.17)(6.132,4.000){2}{\rule{0.450pt}{0.400pt}}
\multiput(851.00,516.60)(1.066,0.468){5}{\rule{0.900pt}{0.113pt}}
\multiput(851.00,515.17)(6.132,4.000){2}{\rule{0.450pt}{0.400pt}}
\multiput(859.00,520.60)(0.920,0.468){5}{\rule{0.800pt}{0.113pt}}
\multiput(859.00,519.17)(5.340,4.000){2}{\rule{0.400pt}{0.400pt}}
\multiput(866.00,524.60)(1.066,0.468){5}{\rule{0.900pt}{0.113pt}}
\multiput(866.00,523.17)(6.132,4.000){2}{\rule{0.450pt}{0.400pt}}
\multiput(874.00,528.60)(1.066,0.468){5}{\rule{0.900pt}{0.113pt}}
\multiput(874.00,527.17)(6.132,4.000){2}{\rule{0.450pt}{0.400pt}}
\multiput(882.00,532.60)(1.066,0.468){5}{\rule{0.900pt}{0.113pt}}
\multiput(882.00,531.17)(6.132,4.000){2}{\rule{0.450pt}{0.400pt}}
\multiput(890.00,536.60)(1.066,0.468){5}{\rule{0.900pt}{0.113pt}}
\multiput(890.00,535.17)(6.132,4.000){2}{\rule{0.450pt}{0.400pt}}
\multiput(898.00,540.60)(0.920,0.468){5}{\rule{0.800pt}{0.113pt}}
\multiput(898.00,539.17)(5.340,4.000){2}{\rule{0.400pt}{0.400pt}}
\multiput(905.00,544.60)(1.066,0.468){5}{\rule{0.900pt}{0.113pt}}
\multiput(905.00,543.17)(6.132,4.000){2}{\rule{0.450pt}{0.400pt}}
\multiput(913.00,548.60)(1.066,0.468){5}{\rule{0.900pt}{0.113pt}}
\multiput(913.00,547.17)(6.132,4.000){2}{\rule{0.450pt}{0.400pt}}
\multiput(921.00,552.59)(0.821,0.477){7}{\rule{0.740pt}{0.115pt}}
\multiput(921.00,551.17)(6.464,5.000){2}{\rule{0.370pt}{0.400pt}}
\multiput(929.00,557.60)(0.920,0.468){5}{\rule{0.800pt}{0.113pt}}
\multiput(929.00,556.17)(5.340,4.000){2}{\rule{0.400pt}{0.400pt}}
\multiput(936.00,561.60)(1.066,0.468){5}{\rule{0.900pt}{0.113pt}}
\multiput(936.00,560.17)(6.132,4.000){2}{\rule{0.450pt}{0.400pt}}
\multiput(944.00,565.60)(1.066,0.468){5}{\rule{0.900pt}{0.113pt}}
\multiput(944.00,564.17)(6.132,4.000){2}{\rule{0.450pt}{0.400pt}}}
\end{picture}